\documentclass[12pt]{article}
\usepackage{latexsym}
\usepackage{euscript}

\usepackage{amsmath,amssymb}
\usepackage{xcolor}
\usepackage{textcomp}
\usepackage{color}
\usepackage{graphicx}
\usepackage{amsmath,amssymb,color,graphics,cite,footnote}

\usepackage{epsfig}
\usepackage{epstopdf}
\usepackage{mathtools}
\usepackage{ulem}

\def\be{\begin{equation}}
\def\ee{\end{equation}}
\def\bea{\begin{eqnarray}}
\def\eea{\end{eqnarray}}

\long\def\symbolfootnote[#1]#2{\begingroup%
\def\thefootnote{\fnsymbol{footnote}}\footnote[#1]{#2}\endgroup}

 \large\normalsize

\usepackage{silence}
\usepackage{soul}

\begin{document}
\thispagestyle{empty}
\vspace*{0.2cm}
\begin{center}
{\Large \bf Black Hole Shadow and Light Deflection in Generalized Heisenberg-Euler Nonlinear Electrodynamics}
\\
\vspace{1.0cm}
{\large Beyhan Puli{\c c}e\footnote{beyhan.pulice@istinye.edu.tr}$^{\ast}$, Ali \"Ovg\"un\footnote{ali.ovgun@emu.edu.tr}$^{\dagger}$ and Yosef Verbin\footnote{verbin@openu.ac.il}$^{\ddagger}$}
\\
\vspace{0.5cm}
\small{
$^{\ast}$ Department of Basic Sciences, {\.I}stinye University, 34396, {\.I}stanbul, T\"{u}rkiye\\
$^{\dagger}$ {Physics Department, Eastern Mediterranean University, 99628  via Mersin 10, T\"{u}rkiye}\\
$^{\ddagger}$ Astrophysics Research Center, the Open University of Israel, Raanana 4353701, Israel\\}
\end{center}

\begin{abstract}
We study the optical properties of electrically charged black holes sourced by generalized Heisenberg-Euler nonlinear electrodynamics (NLED) in the second-order formalism. Using the exact parametric form of the static spherically symmetric solution, we analyze the photon sphere, the black hole shadow, and the light deflection angle, and compare the results with those of the Reissner-Nordström and Schwarzschild cases. We find that the photon sphere and shadow radii decrease with increasing charge and increase with black hole mass, while NLED effects become significant mainly near extremality. By confronting the dimensionless shadow radius with the Event Horizon Telescope (EHT) bounds from Sgr A* and M87*, we derive phenomenological upper limits on the charge, with Sgr A* providing the stronger constraint. We further show that, for fixed impact parameter and charge, the deflection angle is systematically larger than in the corresponding Reissner-Nordström spacetime, and that the critical impact parameter is shifted to larger values. Our results show that NLED corrections produce observable modifications in the optical properties of charged black holes, especially close to extremality.

\end{abstract}


\newpage

\tableofcontents

\section{Introduction} 
\setcounter{equation}{0}

Nonlinear electrodynamics (NLED) extends Maxwell’s theory by incorporating electromagnetic self-interactions. Two prominent examples are Born–Infeld theory, originally proposed to render the electrostatic self-energy of a point charge finite, and the Heisenberg–Euler effective Lagrangian, which describes quantum corrections to classical electrodynamics arising from vacuum polarization \cite{BornInfeld1934,EulerKockel1933,HeisenbergEuler1936,Sorokin2022}. When coupled to gravity, NLED modifies the electromagnetic stress–energy tensor and the spacetime geometry it sources, providing a framework for investigating how electromagnetic nonlinearities affect black hole solutions and their physical properties \cite{Novello2000,AyonBeatoGarcia1998,Bronnikov2001}. 

Charged black holes in NLED provide a setting for examining departures from the Reissner–Nordström solution of Einstein–Maxwell theory. Depending on the nonlinear Lagrangian and its parameters, electromagnetic self-interactions can modify the electric field profile, the horizon structure, and the relation between black hole mass and charge \cite{Novello2000,AyonBeatoGarcia1998,Bronnikov2001,Sorokin2022}. These modifications motivate the study of the resulting spacetime geometry and its implications for particle motion and light propagation.

Earlier studies by Yajima and Tamaki \cite{Yajima+Tamaki2000} and Ruffini et al. \cite{RuffiniWuXue2013} investigated the effects of Heisenberg–Euler nonlinearities on charged black hole solutions. Subsequent work examined their geodesic structure, perturbations, thermodynamics, and optical properties \cite{Amaro:2020, Breton:2021, Luo:2022, Magos:2020}, while noncommutative-inspired extensions were also developed \cite{Maceda:2018, Maceda:2020}. In comparing these results, it is important to distinguish between the original formulation in terms of the electromagnetic field strength $F_{\mu\nu}$ and the description in terms of the Plebański dual field $P_{\mu\nu}$. For a given NLED theory, these descriptions are equivalent wherever the corresponding Legendre transformation is invertible. However, a polynomial chosen independently in the dual invariants does not generally correspond to the same polynomial in the original electromagnetic invariants. Moreover, when the Lagrangian or its dual is truncated at a specified order in the nonlinear couplings, agreement is expected only to that order. Comparisons between electrically charged solutions must therefore account for both the defining electromagnetic theory and the approximation used.

The generalized Heisenberg–Euler (GHE) model considered in  \cite{our-PRD} extends the polynomial structure of Heisenberg–Euler electrodynamics by allowing nonlinear terms with higher integer powers of the electromagnetic invariants. It provides a phenomenological framework for studying electromagnetic self-interactions beyond the quadratic-invariant case. Within the original second-order formulation, \cite{our-PRD} obtained exact parametric solutions describing static, spherically symmetric, electrically charged black holes. These solutions provide the background geometries for the present study of null geodesics and their associated circular orbits and deflection.

The motivation for studying these solutions is both theoretical and observational. In static, spherically symmetric spacetimes, unstable circular null orbits and critical impact parameters characterize the boundary between capture and scattering, while deflection angles describe the bending of trajectories over a range of distances from the black hole \cite{Synge1966,Falcke:1999pj,Bozza2002,CunhaHerdeiro2018,shadow-review}. These quantities provide a basis for comparing the GHE geometry with the Schwarzschild and Reissner–Nordström solutions. Observationally, the Event Horizon Telescope images of M87* and Sgr A* have made horizon-scale tests of black hole models possible \cite{Akiyama2019M87,EventHorizonTelescope:2022wkp,EventHorizonTelescope:2022xqj}. Connecting theoretical predictions to these observations requires a consistent description of photon propagation and an account of how the observed emission relates to the underlying shadow.

These considerations motivate a quantitative assessment of how the nonlinear electromagnetic sector affects the null geodesic structure of GHE black holes. We examine the dependence of circular null orbits, critical impact parameters, and deflection angles on the black hole mass, electric charge, and nonlinearity index, using the Reissner–Nordström solution as a reference. Particular attention is given to the near-extremal regime, where departures from the linear electromagnetic case may become more pronounced. Establishing the magnitude and parameter dependence of these departures is necessary before assessing their potential observational significance.

The present analysis uses the exact parametric GHE black hole solution derived in \cite{our-PRD} from the coupled Einstein and nonlinear electromagnetic field equations in the original second-order formulation. This representation retains the full nonlinear dependence specified by the chosen polynomial Lagrangian, without truncating the background solution in the nonlinear coupling. It therefore allows us to examine the null geodesic structure beyond a weak-nonlinearity expansion and to assess the range over which the Reissner–Nordström geometry provides an accurate approximation. The focus here is on the consequences of this exact background for circular null orbits and scattering trajectories.

In this paper, we determine the unstable circular null orbits and critical impact parameters of the GHE background and compute deflection angles for trajectories scattered from infinity back to infinity. We compare these results with the Schwarzschild and Reissner–Nordström cases within their respective black hole parameter domains. We also compare the shadow radius obtained in the background null-geodesic description with published EHT bounds for M87* and Sgr A*. This comparison is interpreted as a conditional phenomenological test of the background geometry; its application to electromagnetic observations requires specifying the photon propagation law. 

The paper is organized as follows. Section $2$ introduces the formalism for null geodesics and shadow formation in static, spherically symmetric spacetimes. Section $3$ reviews the electrically charged GHE black hole solution in the second-order formulation. Section $4$ examines circular null orbits and the associated shadow radius in the background geometry and presents a comparison with published EHT bounds. Section $5$ studies deflection angles and compares the GHE results with the Reissner–Nordström and Schwarzschild cases. Section $6$ summarizes the findings and discusses their interpretation.

\section{Brief Review of Photon Motion around Black Holes}
\setcounter{equation}{0}

\subsection{Photon Motion}

In this section, we study the photon motion around a static spherically symmetric (SSS)
black hole (BH) with line element is
\begin{equation}d s^{2}= f(r) d t^{2}-\frac{1}{f(r)} d r^{2}-r^{2}\left(d \theta^{2}+\sin ^{2} \theta d \phi^{2}\right) \, 
\label{line-element}
\end{equation}
where $f(r)$ is the metric function. We also assume asymptotic flatness, namely $f(r)\rightarrow 1$ as  $r\rightarrow \infty$ so that the geometry approaches Minkowski spacetime at large radial distances. To analyze the motion of test particles in this background, we employ the Lagrangian formalism for geodesics. The geodesic equations follow from the Lagrangian
\begin{align}
\label{eq-lag}
\mathcal{L} = - \frac{1}{2}g_{\mu \nu} \dot{x}^\mu \dot{x}^\nu \,
\end{align}
where a dot denotes differentiation with respect to an affine parameter $\lambda$. For the SSS metric \eqref{line-element}, this becomes
\begin{equation}
\label{Lagrangian}
   \mathcal{L} = - \frac{1}{2} \left ( f(r)\dot{t}^2 - \frac{\dot{r}^2}{f(r)}- r^2 \dot{\theta}^2  - r^2 \sin^2\theta \dot{\phi}^2 \right ) .
\end{equation}
Spherical symmetry allows the orbital plane to be chosen as $\theta = \pi/2$, with $\dot{\theta}=0$, without loss of generality. Since $t$ and $\phi$ are cyclic coordinates, their conjugate momenta are conserved. With the sign convention adopted in \eqref{eq-lag}, we define
\begin{equation} \label{energy}
    E = f(r)\frac{dt}{d\lambda},
\end{equation}
and
\begin{equation} \label{ang-mom}
    l = r^2 \frac{d\phi}{d\lambda}.
\end{equation}
The constants $E$ and $l$ are the conserved quantities, energy and angular momentum, associated with time translations and rotations about the axis normal to the orbital plane, respectively.

A third conserved quantity is the Lagrangian \eqref{Lagrangian} itself whose conserved value we denote as $\mathcal{L}=-\frac{1}{2} \epsilon$ where the parameter $\epsilon$ distinguishes the causal nature of the trajectory. For timelike geodesics, corresponding to massive particles, one has $\epsilon = 1$, and the affine parameter may be identified with the proper time. For null geodesics, corresponding to photons, one has $\epsilon = 0$, and the trajectory is parametrized by an arbitrary affine parameter.

Substituting the two conservation laws (\ref{energy}) and (\ref{ang-mom}) into (\ref{Lagrangian}) together with $\mathcal{L}=-\epsilon/2$, we obtain the first-order radial equation 
\begin{align}
\label{eq-energy-cons}
\dot{r}^2 + f(r) \left( \frac{l^2}{r^2} + \epsilon \right) = E^2 \; .   
\end{align}
This equation has the form of an energy conservation law for one-dimensional motion in an effective potential. Equation \eqref{eq-energy-cons} defines the radial effective potential
\begin{equation}
    V_{eff}(r) = f(r) \left( \frac{l^2}{r^2} + \epsilon \right).
    \label{eq-eff-pot-dynamics}
\end{equation}
Since our main interest is in photon motion, we restrict ourselves to null geodesics by setting $\epsilon = 0$, in which case
\begin{align}
 V^{(0)}_{eff}(r) = \frac{l^2 f(r)}{r^2}  \,.
    \label{eq-eff-pot-photon}    
\end{align}
An alternative and particularly useful description to analyze geodesic motion is to describe the trajectory in its geometric form, namely through the $\phi$-dependent orbit $r(\phi)$. This representation is especially convenient for the study of photon bending and the characterization of critical orbits. Using the conserved angular momentum relation \eqref{ang-mom}, one can rewrite the radial equation \eqref{eq-energy-cons} by changing the affine parameter from $\lambda$ to the azimuthal angle $\phi$. In this way, the geodesic equation can be cast into the form 
\begin{align}
\label{eq-Phi-Dept-trajectory} 
\left ( \frac{dr}{d\phi}  \right)^2 + W_{eff}(r) = 0
\end{align}
where
\begin{align}
 \label{eq-eff-pot-Phi-Dept}
W_{eff}(r) =   \frac{r^4}{\ell^2}  \left ( V_{eff}(r)-E^2\right ) .  \end{align}
Notice  that \eqref{eq-Phi-Dept-trajectory} with (\ref{eq-eff-pot-Phi-Dept})   has the structure of an energy conservation law in one-dimensional classical mechanics with vanishing ``effective energy''. For null geodesics, this expression simplifies considerably. Setting $\epsilon = 0$ and introducing the impact parameter $b=l/E$, \eqref{eq-eff-pot-Phi-Dept} then reduces to
\begin{align}
 \label{eq-eff-pot-geometry}
W^{(0)}_{eff}(r)=  r^2 f(r)-\frac{r^4}{b^2} =  r^2 f(r) \left ( 1- \frac{\gamma^2(r)}{b^2} \right )   \end{align}
where we have defined 
\begin{align}
\label{gamma-func}
\gamma^2(r) = \frac{r^2}{f(r)} \, .
\end{align}

\subsection{Black Hole Shadow}

We next consider circular null geodesics of the background metric in the static exterior. A central role in this analysis is played by the photon sphere, namely the radius of the unstable circular null orbit around a SSS BH. The existence of this orbit can be inferred from the special class of solutions for which the radial coordinate remains constant along the trajectory, that is, $r(\phi)$. 

Such a solution exists when both the first and second derivatives of $r(\phi)$ vanish. The first condition,
\begin{align}
\frac{dr}{d\phi}=0 \, .  
\end{align}
The orbit equation \eqref{eq-Phi-Dept-trajectory} therefore requires
\begin{align}
\label{cond-orbit-1}
W_{eff}(r)=0 \, .  
\end{align}
Using \eqref{eq-eff-pot-geometry}, this condition gives $\gamma^2(r) = b^2 $. This relation alone also holds at an ordinary radial turning point and is therefore insufficient to establish circular motion. A circular geodesic must additionally satisfy
\begin{align}
\frac{d^2 r}{d\phi^2} =0    
\end{align}
which ensures that the turning point is in fact a circular orbit rather than a generic point of closest approach. The radial geodesic equation then yields the second condition,
\begin{align}
\label{cond-orbit-2}
\frac{d W_{eff}}{dr} = 0 \, .    
\end{align}
The resulting circular null orbit is typically unstable and defines the photon sphere radius, denoted by $r_{ps}$.

Combining \eqref{cond-orbit-1} and \eqref{cond-orbit-2}, we obtain
\begin{align}
\frac{d(\gamma^2(r))}{dr}\bigg|_{r=r_{{ps}}}=0 \, .
\end{align}
Since $\gamma^{2}(r)=r^{2}/f(r)$, the circular-orbit condition can equivalently be written as
\begin{equation}
\label{eq-photon-sphere}
\frac{f^{\prime}\left(r_{p s}\right)}{f\left(r_{p s}\right)}-\frac{2}{r_{p s}}=0\,.
\end{equation}
Here and below, a prime denotes differentiation with respect to $r$. The radius $r_{ps}$ determines the null orbit that separates the captured and scattered null trajectories. Consequently, it also fixes the critical impact parameter and hence the apparent size of the black hole shadow as seen by a distant observer.
 \begin{figure}[b!]
\begin{center}
\includegraphics[width=0.8\textwidth]{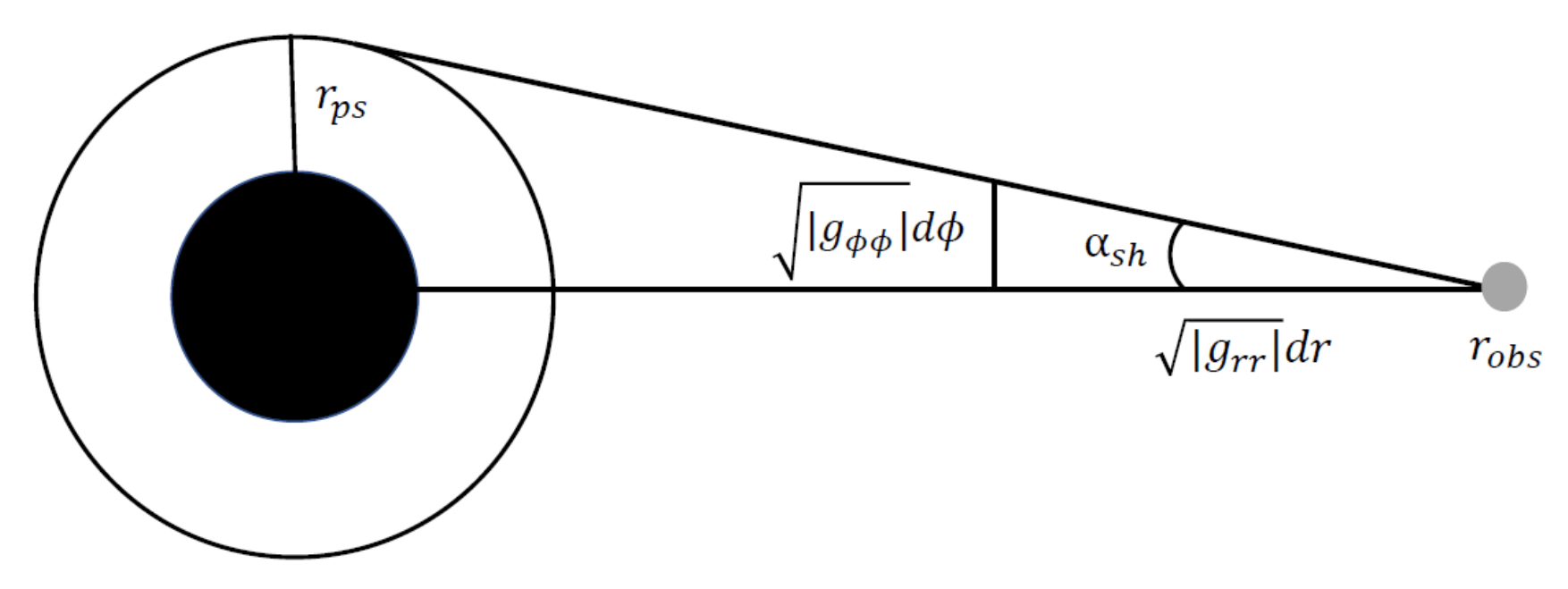}
\caption{\small{Illustration of shadow calculation from the observer’s position $r_{obs}$ (represented by the grey small disk). The shadow angle $\alpha_{sh}$ is the angle between the limiting light ray (which arrives from the photon sphere (radius $r_{ps}$) at the observer) and the radial direction. The infinitesimal angular and radial lengths are calculated at $r=r_{obs}$}.}
\label{shadow_diagram}
\end{center}
\end{figure}

To determine the black hole shadow, we consider a static observer located at radial position $r_{obs}$ and trace light rays backward in time as illustrated in Fig.\ref{shadow_diagram}. The angular position of an incoming photon with respect to the radial direction is characterized by the angle $\alpha$. For the metric given above, this angle satisfies \cite{shadow-review}
\begin{align}
\cot{\alpha} = \sqrt{\frac{\lvert g_{rr}\rvert}{\lvert g_{\phi \phi}\rvert}} \frac{dr}{d\phi} \bigg|_{r=r_{obs}} = \frac{1}{\sqrt{r^2 f(r)}} \frac{dr}{d\phi} \bigg|_{r=r_{obs}}
\end{align}
The boundary of the shadow is determined by the critical null rays that asymptotically approach the photon sphere when traced backward in time. These rays separate photons that fall into the black hole from those that escape to infinity, and therefore define the angular radius of the shadow, denoted by $\alpha_{sh}$. 

Evaluating the geodesic equation \eqref{eq-Phi-Dept-trajectory} together with the null effective potential (\ref{eq-eff-pot-geometry}) for the critical null trajectory at the observer position $r=r_{obs}$ and at the photon sphere $r=r_{ps}$, one obtains
\begin{align}
\cot^2{\alpha_{sh}} = \frac{\gamma^2(r_{obs})}{\gamma^2(r_{ps})} -1 \, .
\end{align}
This immediately yields the angular radius of the black hole shadow in the form
\begin{align}
\label{shadow-angle}
\sin^2 \alpha_{sh}=\frac{\gamma^2(r_{ps})}{\gamma^2(r_{obs})} \, .
\end{align}
Using the definition of $\gamma(r)$ (\ref{gamma-func}), the above expression can be written explicitly as
\begin{eqnarray}\label{shadow-angle-2}
\sin^2 \alpha_{sh}=\frac{r_{ps}^2}{f(r_{ps})}\frac{f(r_{obs})}{r^2_{obs}}.
\end{eqnarray}
For an observer located far from the black hole, the shadow radius can be approximated as
\begin{eqnarray}
\label{shadow-radius}
r_{sh}&\simeq&r_{obs} \sin \alpha_{sh} \simeq \frac{r_{ps}}{\sqrt{f(r_{ps})}}{\sqrt{f(r_{obs})}}.
\end{eqnarray}
In the asymptotically flat limit, where $f(r_{obs})\to 1$ for $r_{obs}\to\infty$, this expression reduces to
\begin{equation}
r_{sh}\simeq \frac{r_{ps}}{\sqrt{f(r_{ps})}}=b_c,
\end{equation}
showing that the apparent shadow radius seen by a distant observer is equal to the critical impact parameter $b_c$.

As a benchmark example, let us consider the RN black hole with metric function
\begin{align}
f(r)=1-\frac{2M}{r}+\frac{Q^2}{2r^2}
\label{RN-metric}\, ,     
\end{align}
where $M$ and $Q$ denote the mass and electric charge of the black hole, respectively. The photon sphere radius is then given by
\begin{align}
\label{eq-RN-photon-sphere}
r_{ps} = \frac{1}{2} \left ( 3 M + \sqrt{9M^2 -4Q^2}  \right ) \, 
\end{align}
which  satisfies $r_{ps} > r_h$ where the horizon radius is
\begin{align}
r_h = M + \sqrt{M^2- \frac{Q^2}{2}} \, .     
\end{align}
The RN black hole domain is therefore $Q^{2}\leq2M^{2}$, with equality corresponding to extremality. All comparisons with RN black holes are restricted to this domain.

Using \eqref{shadow-radius} with $f(r_{obs})\rightarrow 1$, the shadow radius of the RN black hole becomes
\begin{align}
r_{sh} = \frac{r^2_{ps}}{\sqrt{r^2_{ps} - 2 M r_{ps} + \frac{Q^2}{2}}}  \, .
\end{align}
Finally, substituting (\ref{eq-RN-photon-sphere}), one obtains
\begin{align}
r_{sh} = \frac{\left ( 3M +\sqrt{9M^2 - 4Q^2} \right )^2}{2 \sqrt{6M^2-2Q^2+2M\sqrt{9M^2-4Q^2} }} \, .
\end{align}
At fixed mass, the shadow radius decreases monotonically with increasing $|Q|$ over the black hole domain, from $3\sqrt{3} M$ in the Schwarzschild limit to $4M$ at extremality.

In this context, it is convenient to recall the observational bounds on the dimensionless shadow radius 
from EHT measurements. For Sgr A*, the $1 \sigma$ range is given by \cite{shadow-constraints} 
\begin{align}
\label{Sgr-EHT}
4.55 \le r_{sh}/M \le 5.22 \, ,
\end{align}
while for M87*, the corresponding $1 \sigma$ range  reads \cite{EHT-M87}
\begin{align}
\label{M87-EHT}
4.31 \le r_{sh}/M \le 6.08 \, .
\end{align}
These observational bounds will be used below to assess the phenomenological viability of the GHE shadow predictions.

For reference, in the Schwarzschild black hole limit ($Q=0$) one has 
\begin{align}
\frac{r_{\rm ps}}{M} = 3,  \qquad  \frac{r_{\rm sh}}{M} = 3\sqrt{3} \;\approx\; 5.196     
\end{align}
It is worth noting that the Schwarzschild value lies very close to the upper edge of the EHT $1\sigma$ band for Sgr A* \eqref{Sgr-EHT}. This indicates that the current shadow measurement already places nontrivial constraints on charged black hole geometries. In particular, within an RN-type spacetime, the introduction of charge decreases the dimensionless shadow radius relative to its Schwarzschild value. Consequently, even modest departures from the neutral case may induce a noticeable shift of $r_{\rm sh}/M$ within the observationally allowed range. This illustrates the constraining power of present EHT data on charged black hole solutions. This highlights the constraining power of current EHT data on charged black hole solutions.

\section{Brief Review of GHE  Electrostatic Black Holes}
\setcounter{equation}{0}
In this section, we describe SSS electric BH  solutions in the framework of generalized Heisenberg-Euler (GHE) theory in the second-order formalism, following the construction presented in detail in our previous work \cite{our-PRD}. Studying BH solutions arising from NLED theories is an important and compelling direction of research. Our analysis focuses on the GHE theory minimally coupled to GR, for which exact analytical black hole solutions were obtained in \cite{our-PRD}, together with the associated physical quantities such as the mass, temperature, entropy, and related thermodynamic characteristics.

The Lagrangian density of the GHE theory, minimally coupled to gravity is given by
\begin{equation}
\mathcal{L}=\frac{1}{2\kappa}R-\frac{1}{4}F_{\mu\nu}F^{\mu\nu}+\frac{\gamma}{2n}(F_{\mu\nu}F^{\mu\nu})^n +\frac{\beta}{2n}(F_{\mu\nu}\;^{*}F^{\mu\nu})^n\; ,
\label{EqSelf-Grav-NLED-2ndOrder}
\end{equation}
where $\gamma$ and $\beta$ are real parameters. The fundamental dynamical variable of the theory is the  4-potential $A_\mu$ from which the field-strength tensor is defined in the usual way as
\begin{align}
F_{\mu\nu}=\partial_\mu A_\nu-\partial_\nu A_\mu \, , 
\end{align}
while its dual is given by 
\begin{align}
^{*}F^{\mu\nu} = \epsilon^{\mu\nu\rho\sigma}F_{\rho\sigma}/2\sqrt{-g} \, .
\end{align} 

Varying the action corresponding to the above Lagrangian \eqref{EqSelf-Grav-NLED-2ndOrder} with respect to the metric yields the Einstein field equations,
\begin{align}
G_{\mu\nu}=-\kappa T_{\mu\nu}    
\end{align}
where the associated energy-momentum tensor takes the form
\begin{align}
T_{\mu\nu}=&-\left(1- 2\gamma   (F_{\sigma\beta}F^{\sigma\beta}) ^{n-1} \right)F_{\mu\alpha}F_{\nu}^{\ \alpha} \nonumber \\
&+ \frac{1}{4}\left( F_{\sigma\beta}F^{\sigma\beta} - \frac{2\gamma}{n}  (F_{\sigma\beta}F^{\sigma\beta})^n + \frac{2(n-1)\beta}{n} (F_{\sigma\beta}\;^{*}F^{\sigma\beta})^n\right)g_{\mu\nu}
 \ .
\label{TmunuGEH}
\end{align}
To construct SSS solutions, we consider a purely radial electric field $F_{tr}(r)$ together with an SSS spacetime. Under these assumptions, the energy-momentum tensor satisfies $T_t^t=T_r^r$ and $T_\theta^\theta=T_\phi^\phi$, which is consistent with the metric form given in \eqref{line-element}. As a result, there remain only two independent components of $T_\mu^\nu$, and correspondingly only two algebraically independent Einstein equations. These equations are not independent dynamically, however, since they are related through the Bianchi identities. Therefore, it is sufficient to solve a single Einstein equation in order to determine the metric function $f(r)$.

For convenience, we choose the first-order $(tt)$ equation, which takes the form
\begin{eqnarray}
\frac{1}{r^2 }\frac{d}{dr}r(1-f)&=&\kappa \left(\frac{1}{2}F_{tr}^2+\frac{(2n-1)2^{n} |\gamma|}{2n} F_{tr}^{2n} \right) \nonumber \\
\quad \Rightarrow \quad
\frac{dM}{dr} &=& \frac{\kappa \mathfrak{E}^{2} r^2 }{2} \left( \frac{1}{2}\mathcal{E}^2+\frac{(2n-1)}{2n}\mathcal{E}^{2n} \right) \;,
\label{EinsteinEq00FFn2ndOrder}
\end{eqnarray}
where the mass function is introduced through the standard parametrization
\begin{align}
f(r)=1-2M(r)/r \, .
\end{align}
The self-interaction parameter $\gamma$ is traded for an electric field parameter $\mathfrak{E}$ defined by
$(-2)^n \gamma =1/\mathfrak{E}^{2(n-1)}$, such that a dimensionless electric field is defined as $\mathcal{E}=F_{tr}/\mathfrak{E}$ keeping the additional condition $(-1)^n\gamma >0$ which is needed for a well-behaved electric field that requires the sign of  $\gamma$ to be correlated with the power $n$ \cite{our-PRD}.

The Coulomb law for $F_{tr}$ is easily integrated to the following dimensionless equation
\begin{equation}
\mathcal{E}+\mathcal{E}^{2n-1} = q/\varrho^2
\label{EqFtrSecondOrderGen_nDmlss},
\end{equation}
 where we have introduced the length scale $\ell = 1/\sqrt{\kappa \mathfrak{E}^{2}}$, together with the dimensionless radial coordinate $\varrho=r/\ell$ and charge parameter $\varrho=r/\ell$ and $q = \kappa \mathfrak{E}\, Q$. We will also use the dimensionless mass function $m=M/\ell$ so that the Einstein equation \eqref{EinsteinEq00FFn2ndOrder} can likewise be written in dimensionless form as
\begin{equation}
1-f(\varrho) -\varrho \frac{df}{d\varrho}  = \varrho^2 u(\mathcal{E}) \;\;\; , \;\;\;
\frac{dm}{d\mathcal{E}} = \frac{1 }{2} \varrho^{2}(\mathcal{E})u(\mathcal{E})\frac{d\varrho}{d\mathcal{E}}
\label{EqStaticSphericalMetric2}
\end{equation}
where 
\begin{align}
u(\mathcal{E})=\frac{1}{2}\mathcal{E}^2+\frac{(2n-1)}{2n} \mathcal{E}^{2n}    
\end{align}
is the dimensionless energy density.

Integrating the second equation of \eqref{EqStaticSphericalMetric2}, one finds  the mass function in terms of Gauss hypergeometric functions $F(a,b, c, z)$:
\begin{align}
m(\mathcal{E})&= q^{3/2} \left(\frac{ 5 n+ (6 n-1) \mathcal{E}^{2
   (n-1)}}{12 n \left(1+ \mathcal{E}^{2(n-1)}\right)^{3/2}}\mathcal{E}^{1/2}
 \hspace{4.0cm} \right. \nonumber \\ 
&\left. +\frac{2 }{3 (2 n-3) \, \mathcal{E}^{n-\frac{3}{2}}}  \,
  F\left(\frac{1}{2},\frac{2 n-3}{4 (n-1)},\frac{6 n-7}{4 (n-1)},
  -\frac{1}{ \mathcal{E}^{2 (n-1)}}\right)  \right) \nonumber \\ 
&+ m_{BH} - \frac{2q^{3/2}}{3 \sqrt{\pi}} \frac{\Gamma\left(\frac{1}{4 (n-1)}\right) \Gamma \left(\frac{6 n-7}{4(n-1)}\right)}{2n-3 } \, .
   \label{MassFunctionNLED2ndOrder}
\end{align}   
Using Eqs. \eqref{EqFtrSecondOrderGen_nDmlss} and \eqref{MassFunctionNLED2ndOrder}, the metric function $f(r)$ can be expressed analytically in parametric form as
\vspace{0.05cm}
\begin{eqnarray}
\left\{
\begin{array}{rl}
&f(\mathcal{E})= 1-\frac{\displaystyle{2 }}{ \displaystyle{q^{1/2} }}\left(\mathcal{E}+\mathcal{E}^{2n-1}\right)^{1/2} m(\mathcal{E})  \vspace{0.2cm}\\ 
&r(\mathcal{E})= \ell \cdot \varrho(\mathcal{E}) =  \frac{ \displaystyle{ \ell q^{1/2}}}{\displaystyle{\left(\mathcal{E}+\mathcal{E}^{2n-1}\right)^{1/2}}}  \, \, \, . \\
\end{array} \right.
   \label{MetricFctnNLED2ndOrder}
\end{eqnarray}
With the analytic parametric form of the spacetime geometry at hand, we may now turn to its observational implications. Our next goal is to determine the photon sphere structure and the associated black hole shadow, and to compare the resulting predictions with the EHT constraints.

\section{Shadow Cast with EHT Constraints}
\setcounter{equation}{0}
In this section, we investigate the shadow properties of the GHE black holes described by the spacetime ($\ref{line-element}$), with the metric function given in the parametric form (\ref{MetricFctnNLED2ndOrder}). By confronting the resulting shadow predictions with the EHT observations of Sgr. A* and M87*, we obtain constraints on the charge of the GHE black hole.

We begin by determining the photon sphere radius from the parametric representation of the metric function. But we begin by a simple condition which is most easily found from Eq.\eqref{eq-photon-sphere} (in dimensionless form) combined with the first equation of \eqref{EqStaticSphericalMetric2} which yield together 
\begin{equation}
1-3f(\varrho_{ps})=\varrho^2_{ps}u(\varrho_{ps})
\label{EqPhotonSphereAlgebraic}
\end{equation}
where $\varrho_{ps}$ denotes the dimensionless photon sphere radius.
In our case where parametric representation is used it is simpler to rewrite this equation in terms of $m(\mathcal{E}_{ps})$, namely:
\begin{equation}
 \frac{6 m(\mathcal{E}_{ps})}{\varrho (\mathcal{E}_{ps}) }-2  =\varrho^2 (\mathcal{E}_{ps}) u(\mathcal{E}_{ps}) .
 \label{eq-PS-parametric}
\end{equation}
Substituting the explicit expressions for $m(\mathcal{E})$ from (\ref{MassFunctionNLED2ndOrder}) and $\rho(\mathcal{E})$ from (\ref{MetricFctnNLED2ndOrder}), we arrive at the following explicit equation in terms of $\mathcal{E}_{ps}$
\begin{align}
\label{eq-PS-parametric-2}
0 = & ~ 3 m_{BH} - \frac{\sqrt{q} (1-q \mathcal{E}_{ps} )}{\sqrt{\mathcal{E}_{ps}+\mathcal{E}_{ps}^{2n-1}}} 
- \frac{2 q^{3/2}}{ (2n-3) \sqrt{\pi}} \left [  \Gamma \left ( \frac{1}{4(n-1)}\right ) \Gamma \left ( \frac{6n-7}{4(n-1)}\right )  \right.  \nonumber \\
&\left. - \mathcal{E}_{ps}^{\frac{3}{2}-n} \sqrt{\pi} F \left ( \frac{1}{2}, \frac{2n-3}{4(n-1)},\frac{6n-7}{4(n-1)},-\frac{1}{\mathcal{E}_{ps}^{2n-2}}
\right )  \right ] .
\end{align}
For fixed values of the BH mass $m_{BH}$ and the charge $q$, Eq.~\eqref{eq-PS-parametric-2} can be solved numerically to determine the field value $\mathcal{E}_{ps}$ at the photon sphere. The corresponding photon sphere radius $\varrho_{ps}$ then follows directly from the parametric relation \eqref{MetricFctnNLED2ndOrder}.

\begin{figure}[ht!]
\begin{center}
\includegraphics[width=0.485\textwidth]{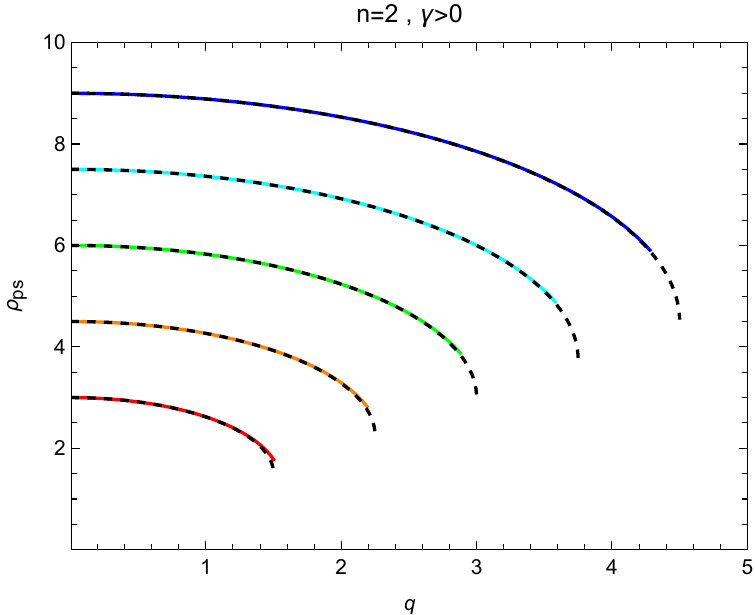}
\includegraphics[width=0.48\textwidth]{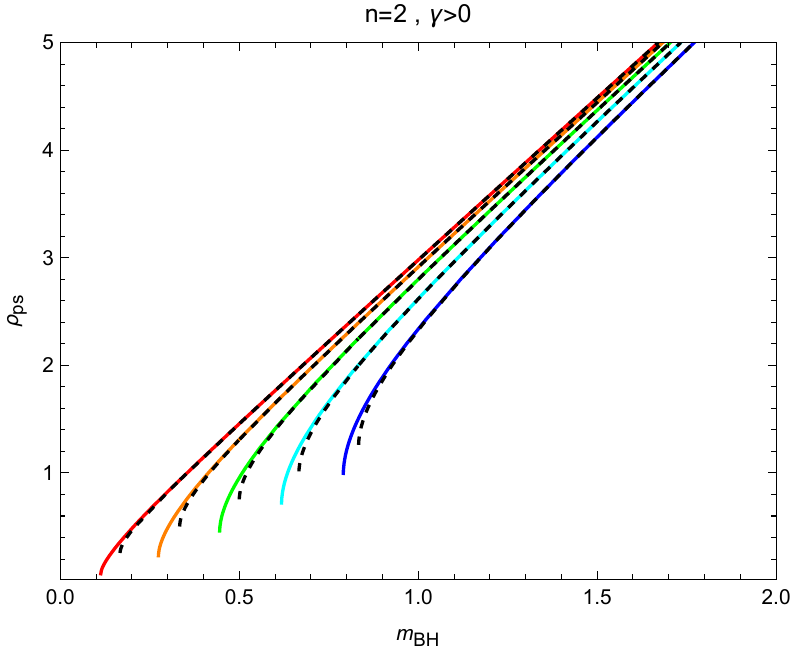}
\caption{\small Photon sphere radius $\varrho_{ps}$ with respect to charge $q$ for $m_{BH} = 1, 1.5, 2, 2.5, 3$ (on the left) and with respect to BH mass $m_{BH}$ for $q = 0.25, 0.5, 0.75, 1, 1.25$ (on the right). The curves of the photon sphere radius have been depicted up to the extremal BH threshold. The black dashed curves represent the RN black hole.}
\label{fig-PS-q-mBH-RN}
\end{center}
\end{figure}

The numerical solution of Eq.~\eqref{eq-PS-parametric-2} allows us to construct Fig.~\ref{fig-PS-q-mBH-RN}, which displays the photon-sphere radius $\varrho_{ps}$ of the GHE black hole as a function of the charge parameter $q$ for fixed values of the black-hole mass $m_{BH}$ (left panel), and as a function of $m_{BH}$ for fixed values of $q$ (right panel). Throughout the figure, a spectroscopic color ordering is adopted, such that the color frequency increases with $m_{BH}$ in the left panel and with $q$ in the right panel. The same convention will be used in the subsequent figures whenever appropriate.

This plot explores the geometric structure of the GHE electrostatic BH by analyzing the photon sphere radius as functions of electric charge and the BH mass. The photon sphere radius decreases monotonically with increasing electric charge for each fixed BH mass (left panel). The curves of the photon sphere radius have been depicted up to the extremal BH threshold since we do not consider naked singularity solutions. Comparing different charge values at a fixed mass reveals that higher charges lead to smaller photon sphere radii. This reflects the fact that increasing the electric charge introduces a repulsive contribution to the spacetime geometry, effectively reducing the gravitational pull on null geodesics and causing the photon sphere to move inward. The photon sphere radius increases monotonically with the BH mass (right panel). This is expected: A more massive BH generates a stronger gravitational field, allowing photons to orbit at larger radii. For a given charge, each curve starts at a finite value of $m_{BH}$, marking the extremal BH threshold.

Fig.~\ref{fig-PS-q-mBH-RN} also provides a direct comparison between the photon sphere radius $\varrho_{ps}$ of GHE black holes (solid curves) and that of Reissner-Nordstr\"om black holes (black dashed curves). In both geometries, the photon sphere radius decreases as the charge increases, reflecting the weakening of null-geodesic trapping induced by the repulsive electric contribution. The agreement between the GHE and RN predictions is very close in the weakly and moderately charged regime. However, noticeable deviations arise as one approaches extremality. In particular, the GHE branches terminate at smaller values of $q$ than the corresponding RN ones, indicating that the maximal charge compatible with a black hole solution is reduced by the NLED corrections. This shows that the GHE solutions reach extremality earlier than their RN counterparts. Moreover, in the near-extremal regime, the GHE photon sphere radius is slightly larger than in the RN case, indicating that the corresponding light-trapping surface is displaced outward.

In the right panel, as expected, the photon sphere radius increases with the black hole mass $m_{BH}$ for both GHE and RN solutions. Over a broad range of masses, the two predictions remain nearly degenerate, indicating that the NLED corrections have only a modest impact in this regime. Noticeable deviations emerge only as one approaches the extremal threshold, where the GHE curves become slightly shifted relative to their RN counterparts. 

This behavior shows that the effect of the nonlinear corrections on the photon sphere structure remains small for weakly charged black holes, but becomes increasingly relevant in the strong-field, near-extremal regime.

\begin{figure}[b!]
\begin{center}
\includegraphics[width=0.6\textwidth]{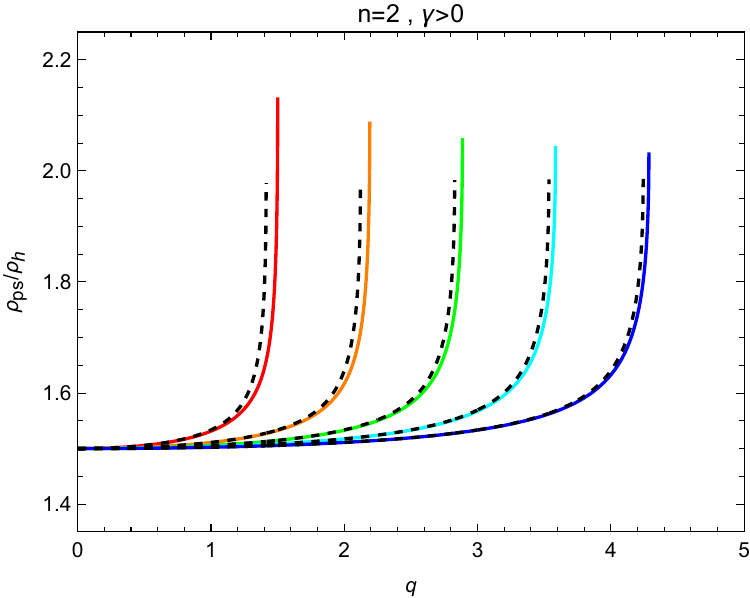}
\caption{\small The ratio of the photon sphere radius to the horizon radius $\varrho_{ps}/\varrho_{h}$ with respect to charge $q$ for $m_{BH} = 1, 1.5, 2, 2.5, 3$. The black dashed curves represent the RN black hole.}
\label{fig-PS-ratio-RN}
\end{center}
\end{figure}

Fig.~\ref{fig-PS-ratio-RN} displays the ratio of the photon sphere radius to the horizon radius, $\varrho_{ps}/\varrho_h$, as a function of the charge $q$ for several values of the black hole mass $m_{BH}$ within the GHE model for $n=2$ and $\gamma>0$. The corresponding RN results are shown by the black dashed curves for comparison.

For all masses considered, the ratio starts from the Schwarzschild value at $q=0$ and increases monotonically with the charge. This shows that, although both the horizon radius and the photon sphere radius decrease as the charge increases, the horizon shrinks more rapidly, so that the photon sphere becomes progressively more separated from the horizon in relative terms. The sharp rise of the curves near their endpoints signals the approach to the extremal regime, where this hierarchy becomes particularly pronounced.

The comparison with the RN case indicates that the two models remain very close throughout the weakly and moderately charged regime, implying that the nonlinear electrodynamic corrections are small in this domain. Noticeable deviations appear only near extremality, where the GHE curves lie slightly above their RN counterparts, especially for smaller values of $m_{BH}$. This behavior shows that the nonlinear corrections enhance the relative separation between the photon sphere and the horizon in the strong-field regime.
 
From a geometrical point of view, the ratio $\varrho_{ps}/\varrho_h$ quantifies how far the photon-trapping surface is located from the horizon in relative terms. It therefore provides a useful characterization of the optical structure of the spacetime, complementary to the direct analysis of the shadow radius presented below. 

Additionally, when examining the behavior of the photon sphere radius, we observed that there is a very weak dependence on $n$. Accordingly, the $n=2$ case is shown here as a representative example.

\begin{figure}[b!!]
\begin{center}
\includegraphics[width=0.475\textwidth]{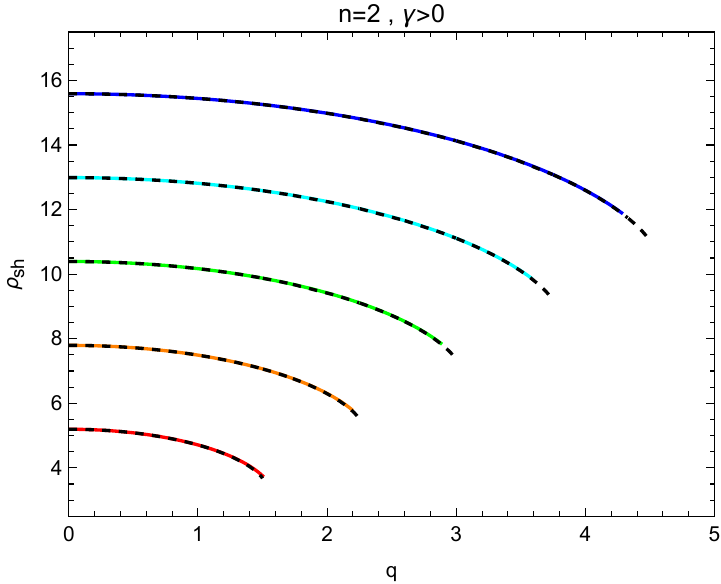}
\includegraphics[width=0.49\textwidth]{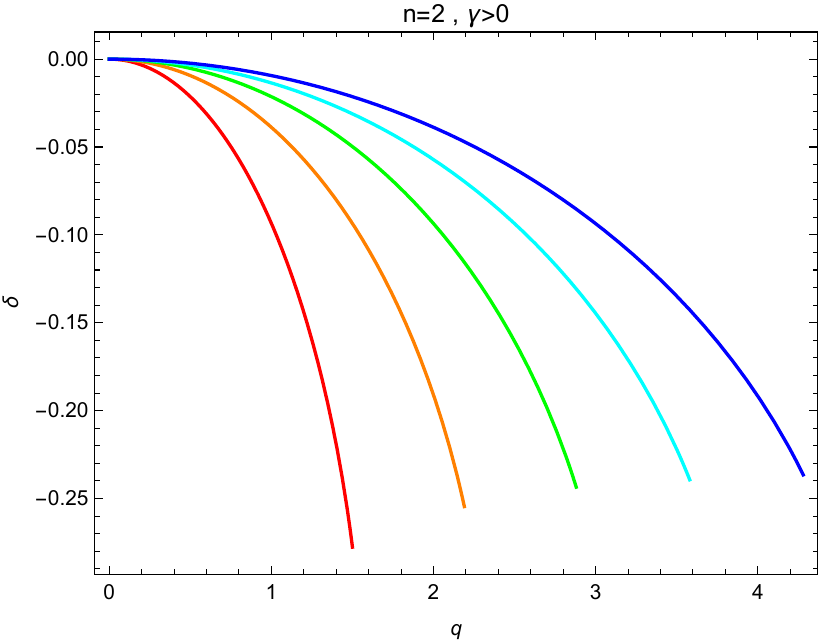}
\caption{\small Shadow radius $\varrho_{sh}$ (on the left) and shadow deviation parameter $\delta$ as a function of the charge $q$ both for $m_{BH} = 1, 1.5, 2, 2.5, 3$ in the GHE theory with $n=2$, $\gamma>0$. The figure depicts each curve extending to the black hole's extremal threshold. The deviation is always negative, indicating that GHE black holes cast smaller shadows than their Schwarzschild counterparts. The black dashed curves represent the RN black hole. }
\label{fig-SH-q-RN}
\end{center}
\end{figure}

Having determined the photon sphere radius, we can now compute the corresponding shadow radius. In the asymptotically flat limit $f(r_{obs}) \rightarrow 1$, (\ref{shadow-radius}) together with the photon sphere relation \eqref{EqPhotonSphereAlgebraic} yields
\begin{equation}
\label{shadow-analytic-Explicit}
\varrho_{sh} =\frac{\sqrt{3}\varrho(\mathcal{E}_{ps})}{\sqrt{1-\varrho^2 (\mathcal{E}_{ps}) u(\mathcal{E}_{ps})}} =\frac{\sqrt{3 q}}{\sqrt{\mathcal{E}_{ps}+\mathcal{E}_{ps}^{2n-1} - q (\frac{1}{2}\mathcal{E}_{ps}^2  + \frac{(2n-1)}{2n} \mathcal{E}_{ps}^{2n}  )}}
\end{equation}
In the second equality, we have used the explicit expressions for $\varrho(\mathcal{E}_{ps})$  and $u(\mathcal{E}_{ps})$.

To quantify the departure of the GHE shadow from the Schwarzschild case, we introduce the Schwarzschild shadow deviation parameter $\delta$, defined as the fractional difference between the shadow radius $\varrho_{sh}$ predicted by the model and the Schwarzschild value $3\sqrt{3}\,M$. Following Refs.~\cite{EHT-M87,EventHorizonTelescope:2022wkp,EventHorizonTelescope:2022xqj,shadow-constraints}, we define
\begin{equation}
\label{eq:shadow_deviation}
\delta \;=\; \frac{\varrho_{sh}}{3\sqrt{3}\,m_{BH}} \;-\; 1 \,.
\end{equation}
This parameter provides a convenient measure of the relative deviation of the GHE shadow from the Schwarzschild prediction. A positive value $(\delta > 0)$, indicates that the GHE black hole casts a larger shadow than a Schwarzschild black hole of the same mass, whereas a negative value $(\delta < 0)$, corresponds to a smaller shadow.

Fig. \ref{fig-SH-q-RN} (on the left) displays the shadow radius $\varrho_{sh}$ as a function of the charge $q$ for several fixed values of the black hole mass $m_{BH}$ in the GHE model with $n=2$. The corresponding RN results are shown by the black dashed curves. For all masses considered, the shadow radius decreases monotonically as the charge increases, showing that the electric sector progressively suppresses the size of the shadow as the solution departs from the Schwarzschild limit. At fixed charge, larger values of $m_{BH}$ correspond to larger shadow radii, as expected from the stronger gravitational field of more massive black holes. In each case, the corresponding branch is shown only up to the extremal black hole threshold.

A direct comparison with the RN geometry shows that the two predictions remain nearly indistinguishable throughout the weakly and moderately charged regime, where the NLED corrections have only a minor effect on the shadow radius. Noticeable deviations emerge only close to extremality. In particular, near the endpoint of each branch, the GHE shadow radius remains slightly larger than its RN counterpart, indicating that the nonlinear corrections become relevant in the strong field region and modify the optical structure of the spacetime. At the same time, the GHE branches terminate at smaller values of $q$ than the RN ones, showing that the maximal charge compatible with a black hole solution is reduced in the presence of NLED effects. Thus, for fixed $m_{BH}$, the GHE solution reaches extremality earlier while exhibiting a milder suppression of the shadow radius in the near-extremal regime.

This slight deviations near extremality suggest that precise shadow measurements, especially for highly charged black holes, could in principle distinguish between GHE and RN models. In this regard, the extremal region serves as a promising window into the nonlinear structure of the spacetimes.

Fig.~\ref{fig-SH-q-RN} (on the right) shows the Schwarzschild shadow deviation parameter $\delta$ as a function of the charge for the same set of masses. A clear and universal trend is observed: $\delta$ remains negative throughout the parameter range considered, and its magnitude increases monotonically with $q$. This implies that, for all configurations shown, the GHE shadow radius is smaller than the Schwarzschild value $3\sqrt{3}M$, with the suppression becoming increasingly pronounced as the black hole approaches extremality.

\begin{figure}[ht!]
\begin{center}
\includegraphics[width=0.6\textwidth]{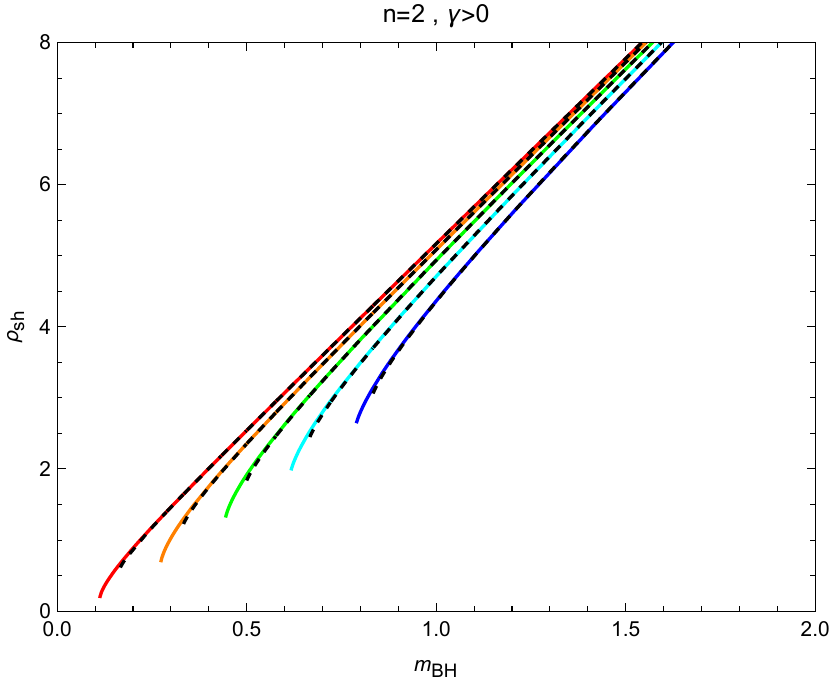}
\caption{\small Shadow radius $\varrho_{sh}$ with respect to black hole mass $m_{BH}$ for $q = 0.25, 0.5, 0.75, 1, 1.25$. The curves show that the shadow radius increases with mass but is suppressed at higher charges. Each curve terminates at the extremal black hole limit. The black dashed curves represent the RN black hole.}
\label{fig-SH-mBH}
\end{center}
\end{figure}

Fig.~\ref{fig-SH-mBH} shows the dependence of the shadow radius $\varrho_{sh}$ on the black hole mass $m_{BH}$ for several fixed values of the charge parameter $q$, both in the GHE model (solid curves) and in the corresponding RN geometry (black dashed curves). In both cases, the shadow radius increases monotonically with $m_{BH}$, in agreement with the expectation that more massive black holes produce larger photon spheres and hence larger shadows.

The GHE and RN predictions remain very close over most of the parameter range shown, indicating that the NLED corrections have only a limited impact on the shadow size away from extremality. Noticeable deviations emerge only near the lower end of each branch, where the solutions approach the extremal threshold. In this regime, the GHE curves are slightly shifted relative to the RN ones, signaling the effect of the nonlinear corrections on the near-horizon optical structure.

Overall, the figure confirms that the shadow radius is primarily controlled by the black hole mass, while the NLED effects remain subleading except in the strong-field, near-extremal regime.

\begin{figure}[b!]
\begin{center}
\includegraphics[width=0.49\textwidth]{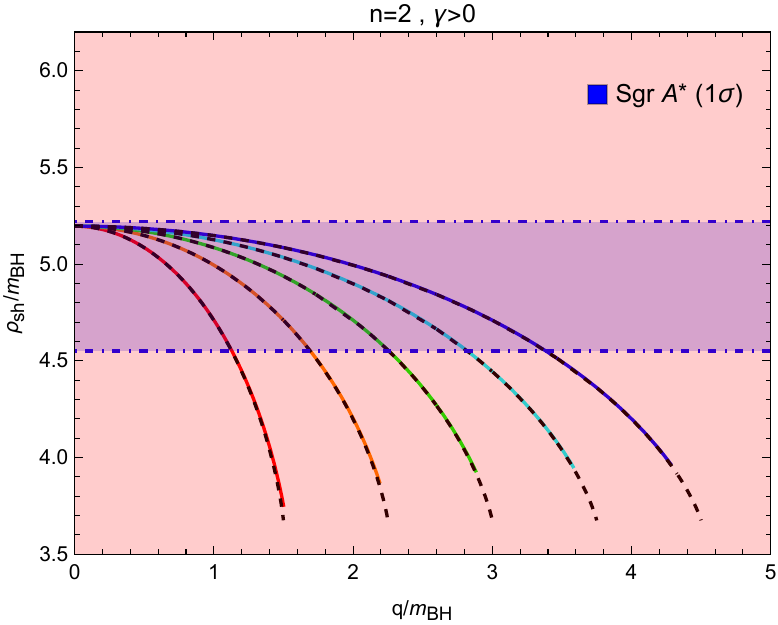}
\includegraphics[width=0.49\textwidth]{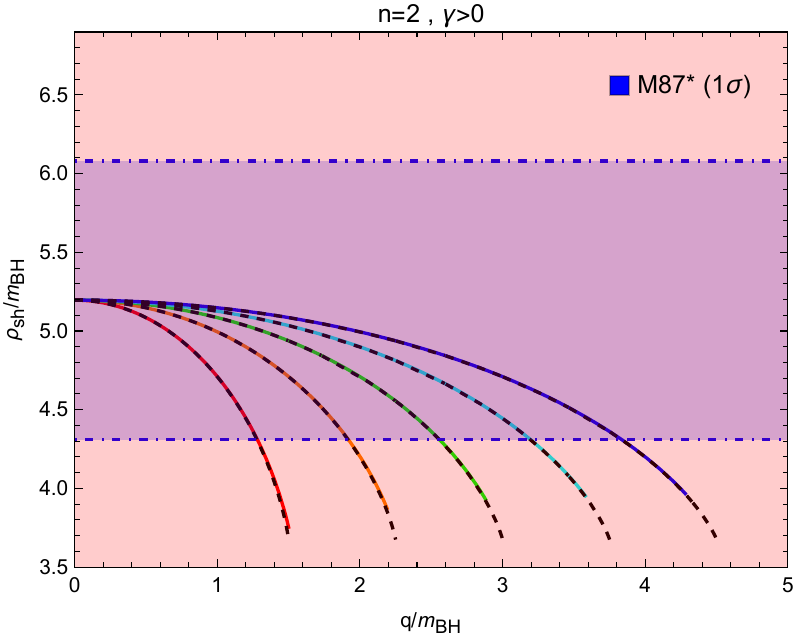}
\caption{\small Shadow radius $\varrho_{sh}/m_{BH}$ with respect to charge $q/m_{BH}$ both in units of BH mass with allowed $1 \sigma$ shadow radius bands for Sgr. A* (left panel) and M87* (right panel) for $m_{BH} = 1, 1.5, 2, 2.5, 3$. The figure depicts each curve extending to the black hole's extremal threshold. The curves are compared against observational bounds from the EHT for Sgr A*  \cite{shadow-constraints} and M87* \cite{EHT-M87}. The black dashed curves represent the RN black hole.}
\label{fig-SH-q-EHT-RN}
\end{center}
\end{figure}

Fig.~\ref{fig-SH-q-EHT-RN} displays the normalized shadow radius $\rho_{sh}/m_{BH}$ as a function of the normalized charge $q/m_{BH}$ for both GHE and RN black holes. The left and right panels correspond to Sgr A* and M87*, respectively. In each panel, the shaded band represents the corresponding $1 \sigma$ observational interval inferred from the EHT measurements. The solid colored curves denote the GHE solutions, while the black dashed curves show the corresponding RN predictions.

In both geometries, the normalized shadow radius decreases monotonically with increasing $q/m_{BH}$. Over the parameter space region allowed by the EHT $1\sigma$ bands, the GHE and RN curves remain nearly degenerate, indicating that the NLED corrections do not lead to an appreciable modification of the shadow size at the level of current observational accuracy. Noticeable differences arise only close to extremality, where the GHE branches terminate earlier than the RN ones because the nonlinear corrections modify the extremality condition and reduce the maximal admissible charge. Although the deviations in the shadow size itself remain small, this earlier termination of the GHE branches constitutes the main qualitative difference between the two models and may become relevant for future higher precision shadow observations.

Since each branch ends at its corresponding extremal charge, the upper edge of the observationally allowed $1\sigma$ band translates into an upper bound on the charge-to-mass ratio $q/m_{BH}$ that remains phenomenologically admissible. For Sgr A*, these maximal allowed values are extracted numerically from the left panel and summarized in Table~\ref{Sgr*-table}. The same analysis is performed for M87*, with the resulting bounds listed in Table~\ref{M87-table}. The comparison shows that the EHT constraint from Sgr A* imposes more stringent upper bounds on $q/m_{BH}$ than the corresponding constraint from M87*.

Overall, the comparison with the EHT shadow measurements of Sgr A* and M87* shows that only a limited range of charge is compatible with the observed shadow size. In this sense, the shadow data provide a direct phenomenological upper bound on the admissible electric charge and thereby constrain the electromagnetic sector of the GHE black hole spacetime.

\begin{table}[ht]
\centering
\begin{tabular}{l c }
\hline\hline
$m_{BH}$ & $q/m_{BH} (1 \sigma$)\\ 
\hline
1 & 1.13  \\ 
1.5 & 1.69   \\ 
2 & 2.25   \\ 
2.5 & 2.82   \\ 
3 & 3.38   \\ 
\hline\hline
\end{tabular}
\caption{Upper bounds on the charge parameter inferred from the EHT $1\sigma$ shadow constraint for Sgr A* \cite{shadow-constraints}, for several values of the black hole mass.}
\label{Sgr*-table}
\end{table}

\begin{table}[ht]
\centering
\begin{tabular}{l c }
\hline\hline
$m_{BH}$ & $q/m_{BH} (1 \sigma$)\\ 
\hline
1 & 1.28   \\ 
1.5 & 1.90   \\ 
2 & 2.55   \\ 
2.5 & 3.19   \\ 
3 & 3.83  \\ 
\hline\hline
\end{tabular}
\caption{Upper bounds on the charge parameter inferred from the EHT $1\sigma$ shadow constraint for M87* \cite{EHT-M87}, for several values of the black hole mass.}
\label{M87-table}
\end{table}

\section{Deflection Angle}
\setcounter{equation}{0}

The deflection angle of a light ray arriving from infinity, scattered by the black hole, and returning to infinity is defined by
\begin{align}
\alpha(\mathcal{E}_0) = \Delta \phi (\mathcal{E}_0) - \pi \, ,
\label{def-angle-expr}
\end{align}
where $\Delta \phi (\mathcal{E}_0)$ denotes the total change in the azimuthal angle along the null trajectory. Here, the turning point value $\mathcal{E}_0$, corresponding to the distance of closest approach of the null geodesic $\rho (\mathcal{E}_0)$, is determined by
\begin{align}
\hat{b}=\frac{\rho (\mathcal{E}_0)}{\sqrt{f(\mathcal{E}_0)}} 
\label{dimensionless-b}
\end{align}
where $\hat{b}=b/\ell$ is the dimensionless impact parameter. Using \eqref{eq-Phi-Dept-trajectory} together with \eqref{eq-eff-pot-geometry}, this quantity can be written in the parametric form 
\begin{align}
\label{delta-geodesic}
\Delta \phi (\mathcal{E}_0) = 2 \int_{\mathcal{E}_0}^{0} \left ( \frac{ \rho^4(\mathcal{E})}{\hat{b}^2} - \rho^2(\mathcal{E}) f(\mathcal{E}) \right ) ^{-1/2}  \rho^\prime(\mathcal{E}) d \mathcal{E} \, .
\end{align}
Substituting \eqref{dimensionless-b} into \eqref{delta-geodesic}, then using \eqref{def-angle-expr}, one obtains the following integral representation for the deflection angle
\begin{align}
\alpha (\mathcal{E}_0) = 2 \int_{\mathcal{E}_0}^{0} d \mathcal{E} \sqrt{q} \left ( \frac{ \rho^4(\mathcal{E})}{\rho^2(\mathcal{E}_0)} f(\mathcal{E}_0) - \rho^2(\mathcal{E}) f(\mathcal{E}) \right ) ^{-1/2} \left ( \frac{-1 + \mathcal{E}^{2(n-1)}(1-2n)}{2(\mathcal{E}+\mathcal{E}^{2n-1})^{3/2}} \right ) - \pi \, .
\label{def-angle-final-form}
\end{align}

\begin{figure}
    \centering
    \includegraphics[width=0.49\linewidth]{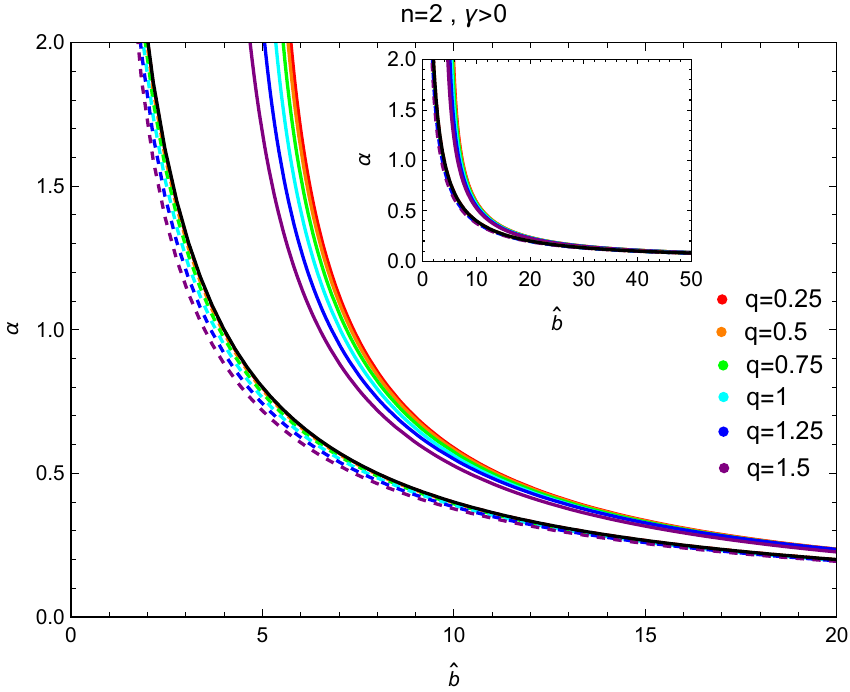}
        \includegraphics[width=0.49\linewidth]{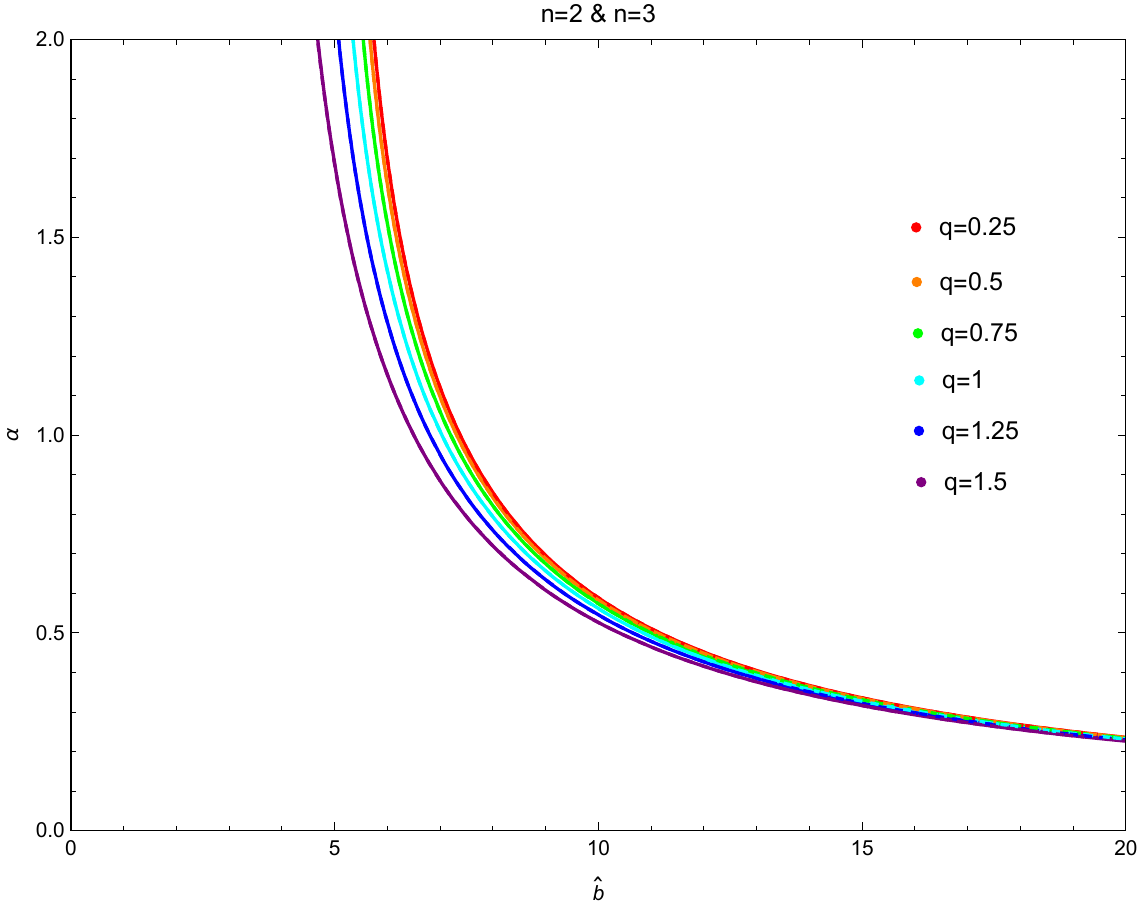}
    \caption{Deflection angle $\alpha$ versus impact parameter $b$ for various charges $q$ in the GHE model (solid) for $n=2$ and $m_{BH}=1$, the RN spacetime (dashed) and the Schwarzschild spacetime $q=0$ (black) (left panel). Comparison between the GHE predictions for $n=2$ (solid) and $n=3$ (dashed) at $m_{BH}=1$ for the same set of $q$ values (right panel).}
    \label{fig:defangle_geodesic_GHE_RN_SCH}
\end{figure}

The integral representation \eqref{def-angle-final-form} cannot be evaluated in closed analytic form, and is therefore treated numerically. For each choice of the parameters, the lower integration limit $\mathcal{E}_0$ is first determined by solving \eqref{dimensionless-b}, after which the deflection angle is computed from \eqref{def-angle-final-form}. The resulting light deflection angle is shown in Fig. \ref{fig:defangle_geodesic_GHE_RN_SCH}. The left panel displays $\alpha$ as a function of the dimensionless impact parameter $\hat{b}$ for the GHE black hole with $n=2$ and $\gamma > 0$, together with the corresponding RN and Schwarzschild cases. The solid colored curves represent the GHE solution for several values of the charge $q$, while the dashed colored curves denote the RN solution for the same charges. The black dashed curve corresponds to the Schwarzschild spacetime with $q=0$.

For all geometries considered, the deflection angle decreases monotonically with increasing impact parameter $\hat{b}$. This reflects the weakening of gravitational bending as the photon trajectory passes farther from the black hole. In the large-$\hat{b}$ regime, all curves approach the Schwarzschild behavior, as expected from asymptotic flatness and the recovery of the standard weak field limit.

A clear distinction nevertheless emerges between the GHE and RN solutions. For fixed values of $(q,\hat{b})$, the deflection angle in the GHE spacetime is systematically larger than in the corresponding RN geometry. This shows that the NLED sector modifies the effective optical geometry experienced by null rays. Hence, although the presence of charge suppresses the bending angle in both models, the nonlinear corrections in the GHE case partially compensate for this suppression and lead to stronger gravitational focusing than in the RN spacetime. In part of the intermediate-$\hat{b}$ regime, the GHE bending angle can even exceed the Schwarzschild value.

These results show that the NLED corrections in the GHE model leave a visible imprint on the light deflection angle. The systematic separation between the GHE, RN, and Schwarzschild curves indicates that light bending is sensitive to the underlying electromagnetic structure of the spacetime,  especially in the strong field regime.

For fixed $\hat{b}$, increasing the charge leads to a systematic decrease in the deflection angle in both the RN and GHE geometries. This behavior reflects the fact that the electromagnetic sector weakens the trapping of null geodesics and hence reduces the bending of light. In the RN case, this suppression is directly associated with the charge contribution in the metric function, which counteracts the purely gravitational attraction. The same qualitative trend persists in the GHE model, although in that case the reduction of the bending angle is modulated by the NLED corrections.

A central feature of the figure is that, for fixed values of $(\hat{b},q)$ , the deflection angle in the GHE spacetime is systematically larger than in the corresponding RN geometry,
\begin{align}
\alpha_{GHE}(\hat{b},q) > \alpha_{RN}(\hat{b},q) \, .   
\end{align}
This shows that the NLED sector in the GHE theory modifies the effective optical geometry in such a way as to enhance the focusing of null rays relative to the standard RN case. Thus, although the presence of charge suppresses the bending angle in both models, the nonlinear structure of the GHE spacetime partially compensates for this suppression, leading to a larger deflection angle than in the linear solution.

The curves terminate at finite values of the impact parameter, corresponding to the charge-dependent critical impact parameter $b_c(q)$ that separates scattering trajectories from captured ones. Fig. \ref{fig:defangle_geodesic_GHE_RN_SCH} shows that this critical value is larger in the GHE spacetime than in the RN case,
\begin{align}
\hat{b}_c^{GHE}(q) > \hat{b}_c^{RN}(q) \, .
\end{align}
This behavior is fully consistent with the corresponding shift of the photon-sphere structure induced by the nonlinear electrodynamic corrections. In particular, it indicates that the threshold between scattering and capture is displaced toward larger critical impact parameters in the GHE geometry relative to the RN solution.

At sufficiently large impact parameters, the GHE and RN curves converge, showing that both geometries recover the Schwarzschild behavior in the far weak field regime. In this limit, the effects of charge and of the NLED corrections enter only at subleading order, in agreement with the standard asymptotically flat deflection behavior.

The right panel of Fig. \ref{fig:defangle_geodesic_GHE_RN_SCH} compares the resulting deflection angle for $n=2$ (solid curves) and $n=3$ (dashed curves), keeping the mass fixed at $m_{BH} = 1$ and considering the same set of charge values $q$. The two families of curves remain nearly degenerate over the entire displayed range of impact parameters, indicating that the deflection angle depends only very weakly on the nonlinearity index in this regime.

\section{Conclusion}
\setcounter{equation}{0}

In this work, we investigated the optical properties of electrically charged black holes sourced by generalized Heisenberg–Euler (GHE) nonlinear electrodynamics (NLED), focusing on the representative case. Using the parametric form of the exact solution, we determined the photon sphere radius, the shadow radius, and the light deflection angle, and compared the resulting predictions with those of the Reissner–Nordström (RN) and Schwarzschild geometries. Our analysis shows that the NLED sector leaves a clear imprint on the null geodesic structure of the spacetime, although these effects remain modest over much of the parameter range and become most pronounced close to extremality.

As shown in Fig. \ref{fig-PS-q-mBH-RN}, the photon sphere radius decreases monotonically with increasing charge and increases with the black hole mass. Relative to the RN solution, the GHE black hole reaches extremality at smaller values of the charge parameter, indicating that the nonlinear electrodynamic corrections reduce the maximal charge compatible with a regular black hole geometry. Fig. \ref{fig-PS-ratio-RN} further shows that the ratio between the photon sphere radius and the horizon radius increases with charge and exhibits a sharper enhancement near extremality, reflecting the fact that the horizon shrinks more rapidly than the photon sphere in this regime. Overall, the GHE and RN photon sphere predictions remain very close away from extremality, while noticeable deviations appear only in the strong field regime.

The shadow analysis leads to a similar conclusion. As illustrated in Figs. \ref{fig-SH-q-RN} and \ref{fig-SH-mBH}, the shadow radius decreases as the charge is increased and grows with the black hole mass, whereas the GHE and RN predictions remain nearly degenerate throughout the weakly and moderately charged regime. Only close to extremality do small but visible deviations emerge, together with the earlier termination of the GHE branches. The Schwarzschild shadow deviation parameter $\delta$, shown in Fig.
\ref{fig-SH-q-RN}, remains negative throughout the parameter range considered and increases in magnitude toward extremality, showing that the GHE shadow is systematically smaller than the Schwarzschild benchmark $3 \sqrt{3} M$. By confronting the dimensionless shadow radius with the $1 \sigma$ constraints from Sgr A* and M87*, as shown in Fig. \ref{fig-SH-q-EHT-RN}, we obtained phenomenological upper bounds on the charge to mass ratio of the GHE black holes. Within the currently allowed observational region, the GHE and RN shadow predictions are almost indistinguishable, so that the main observational difference arises from the fact that the GHE black hole branch terminates earlier and therefore allows a smaller maximal charge-to-mass ratio. In particular, the EHT bounds from Sgr A* were found to impose more stringent constraints on $q / m_{BH}$ than those inferred from M87*.

Our analysis of weak gravitational lensing, summarized in Fig. \ref{fig:defangle_geodesic_GHE_RN_SCH}, further supports this picture. For fixed impact parameter and charge, the GHE deflection angle is systematically larger than in the corresponding RN geometry, indicating that the nonlinear electromagnetic sector modifies the effective optical geometry so as to enhance the focusing of null rays relative to the linear case. In addition, the critical impact parameter is shifted to larger values in the GHE spacetime, consistently with the corresponding displacement of the photon-sphere structure. At large impact parameters, however, the GHE, RN, and Schwarzschild predictions converge, as required by the standard asymptotically flat weak-field limit. We also found that the deflection angle depends only very weakly on the nonlinearity index when comparing the representative cases $n=2$ and $n=3$.

Taken together, these results show that the NLED corrections in the GHE model modify the photon sphere, black hole shadow, and light deflection angle in a mutually consistent way. Although the deviations from the RN geometry are generally modest away from extremality, they become increasingly relevant in the strong field regime, where they alter both the optical structure of the spacetime and the phenomenological bounds inferred from EHT observations.

\vspace{0.4cm}
\noindent{\textbf{Acknowledgments}} 

 B. P. and Y. V. would like to thank for financial support by the Israeli fund for scientific regional cooperation and the Research Authority of the Open University of Israel. A. {\"O}. and B. P. would like to acknowledge the contribution of the COST Action CA21106 - COSMIC WISPers in the Dark Universe: Theory, astrophysics and experiments (CosmicWISPers) and CA23115 - Relativistic Quantum Information (RQI). A. {\"O}., Y. V. and B. P. also thank COST Action CA23130 - Bridging high and low energies in search of quantum gravity (BridgeQG).

\end{document}